\documentclass[twocolumn,showpacs,amsmath,amssymb]{revtex4}
\usepackage{graphicx}
\usepackage{dcolumn}
\usepackage{bm}
\usepackage[dvips]{epsfig}

\newcommand{\ba}{\begin{eqnarray}}
\newcommand{\ea}{\end{eqnarray}}
\newcommand{\ban}{\begin{eqnarray*}}
\newcommand{\ean}{\end{eqnarray*}}
\newcommand{\be}{\begin{equation}}
\newcommand{\ee}{\end{equation}}
\newcommand{\bd}{\begin{displaymath}}
\newcommand{\ed}{\end{displaymath}}

\newcommand{\n}[1]{\label{#1}}

\newcommand{\non}{\nonumber}
\newcommand{\eq}[1]{(\ref{#1})}

\newcommand{\E}{{\mathbb E}}

\newcommand{\hh}{\, ,\hspace{0.5cm}}
\newcommand{\hhh}{\, ,\hspace{0.2cm}}

\newcommand{\pr}[5] {\bibitem{#1} {#2}, Phys. Rev.  {\bf D#3}, #4 (#5).}
\newcommand{\cqg}[5] {\bibitem{#1} {#2},  Class. Quant. Grav.  {\bf #3}, #4 (#5).}
\newcommand{\PRL}[5] {\bibitem{#1} {#2}, Phys. Rev. Lett. {\bf #3}, #4 (#5).}

\begin{document}

\draft


\title{Merger Transitions in Brane--Black-Hole Systems:\\
Criticality, Scaling, and Self-Similarity}
\author{Valeri P. Frolov}
\affiliation{Theoretical Physics Institute, University of Alberta,
Edmonton, Alberta, Canada, T6G 2J1\\
Kavli Institute for Theoretical Physics,  
University of California
Santa Barbara, CA 93106-4030}
\email{frolov@phys.ualberta.ca}

\date{\today}

\begin{abstract}  
We propose a toy model for study merger transitions in a curved
spaceime with an arbitrary number of dimensions. This model includes
a bulk $N$-dimensional static spherically symmetric black hole and a
test $D$-dimensional brane ($D\le N-1$) interacting with the black
hole. The brane is asymptotically flat and allows $O(D-1)$ group of
symmetry.  Such a brane--black-hole (BBH) system has two different
phases. The first one is formed by solutions describing a brane
crossing the horizon of the bulk black hole. In this case the
internal induced geometry of the brane describes $D$-dimensional
black hole.  The other phase consists of solutions for branes which do
not intersect the horizon and the induced geometry does not have a
horizon.  We study a critical solution at the threshold of the
brane-black-hole formation, and the solutions which are close to it.
In particular, we demonstrate, that there exists a striking
similarity of the  merger transition, during which the phase of the
BBH-system is changed, both with the Choptuik critical collapse and
with the merger transitions in the higher dimensional caged
black-hole--black-string system.
\end{abstract}

\pacs{04.70.Bw, 04.50.+h, 04.20.Jb \hfill NSF-KITP-06-14, \ \ \ 
Alberta-Thy-03-06}

\maketitle

\section{Introduction}

Caged  (Kaluza-Klein) black holes, black strings, and possible
transitions between them is a subject which has  recently attracted a
lot of attention (see e.g. reviews \cite{HO,kol1} and references
therein). Kol \cite{kol} demonstrated that during the
black-hole--black-string transition the Euclidean topology of the
system changers. These transitions were called {\em merger}
transitions.  One of the interesting features of the merger
transitions \cite{kol} is their close relation to the Choptuik
critical collapse phenomenon \cite{chop}. The merger transitions are
in many aspects similar to the topology change transitions in the
classical and quantum gravity (see e.g. \cite{hor} and references
therein). In particular, one can expect that during both types of
transitions the spacetime curvature can infinitely grow. It means
that the classical theory of gravity is not sufficient for their
description and a more fundamental theory (such as the string theory)
is required. In these circumstances, it might be helpful to have a
toy model for the merger and topology changing transitions, which is
based on the physics which is well understood. In this paper we would
like to propose such a toy model.

The model consists of a bulk $N$-dimensional black hole and a test
$D$-dimensional brane in it ($D\le N-1$). We assume that the black
hole is spherically symmetric and static. It can be neutral or
charged. As we shall see the detailed characteristics of the black
hole are not important for our purposes. We assume that the brane is
infinitely thin and use the Dirac-Nambu-Goto \cite{D,NG} equations
for its description.  We consider the brane which is  static and
spherically symmetric, so that its worldsheet has the $O(D-1)$ group
of the symmetry. We assume also that the brane reaches the asymptotic
infinity where it has the form of the $(D-1)$-plane. One more
dimension the brane worldsheet gets because of its time evolution. To
keep the brane static one needs to apply a force to it. The detailed
mechanism is not important for our consideration, so that we do not
discuss it here. 

As a result of the gravitational attraction, the brane, which is flat at
infinity, is deformed. There are two different types of possible
equilibrium configurations. A brane either crosses the black hole horizon, or it
lies totally outside the black hole. In the former case the internal
metric of the brane, induced by its embedding, describes a geometry
of $D$-dimensional black hole. This happens because the timelike at
infinity Killing vector of the bulk geometry being restricted to the
brane is the Killing vector for the induced geometry. Thus the brane
spacetime has the Killing horizon (and hence the event horizon) which
is located at the intersection of the brane with the bulk black hole
horizon. A case when the brane lies within the equatorial plane of
the bulk black hole is an example of such a configuration.  Since the
equatorial plane of the spherically symmetric static black hole is
invariant under the reflection mapping the upper half space onto the
lower one, the equatorial plane is a geodesic surface, and hence, it
is a minimal one. Thus the `equatorial' plane automatically satisfies
the Dirac-Nambu-Goto equations. 

An example of a configuration of the second type is  a brane located
at far distance from the black hole. Its geometry is a plane which is
slightly deformed by the gravitational attraction of the bulk black
hole. One can use the weak field approximation to calculate this
deformation \cite{FSS}.

Let us consider a one-parameter family of the branes which are
asymptotically plane and parallel to the chosen equatorial plane. 
This family can be naturally split into two parts (phases). One of
them is formed by sub-critical solutions which do not intersect the
black hole horizon, while the other one is formed by solutions
crossing the horizon. We shall show that there exists a {\em critical}
solution separating these two phases. 

In this paper we study the critical solution at the threshold of the
brane black hole formation, and the solutions which are close to it.
Our goal is to study a transition between the sub- and super-critical
phases. In particular, we demonstrate, that there exists a striking
similarity of this  transition both with the Choptuik
critical collapse \cite{chop} and with the merger transitions in the
black-hole--black-string system \cite{kol,kol1}.

The results presented in the paper are a natural generalization  of
\cite{cfl,flc} adopted to the higher dimensional case \cite{lar}.

\section{Brane equations}

Let us consider a static test brane interacting with a bulk static
spherically symmetrical black hole. For briefness, we shall refer to
such a system (a brane and a black hole) as to the BBH-system.  We
assume that the metric of the bulk $N$-dimensional spacetime is 
\be\n{met}
dS^2=g_{\mu\nu}dx^{\mu}dx^{\nu}=-F dT^2+F^{-1} dr^2 + r^2
d\Omega^2_{N-2}\, ,
\ee
where $F=F(r)$, and $d\Omega^2_{N-2}$ is the metric of
$(N-2)$-dimensional unit sphere $S^{N-2}$.  We define the coordinates
$\theta_i$ ($i=1,\ldots,N-2$) on this sphere by the relations
\be
d\Omega^2_{i+1}= d\theta^2_{i+1} +\sin^2 \theta_{i+1} d\Omega^2_{i}\,
.
\ee

In what follows the explicit form of $F$ is not important. We assume
only that the function $F=F(r)$ has a simple zero at $r=r_0$, where
the horizon of the black hole is located, and it grows monotonically
from $0$  at $r=r_0$ to $1$ at the spatial infinity,  $r\to \infty$
where it has the following asymptotic form
\be\n{as}
F-1\sim r^{-(N-3)}\, .
\ee
For the vacuum (Tangherlini \cite{tang}) solution of the Einstein
equations
\be
F=1-(r_0/r)^{N-3}\, .
\ee

We denote by $X^\mu$ ($\mu=0,\ldots,N-1$)  the bulk spacetime
coordinates and by $\zeta^a\;(a=0,\ldots, D-1)$ the coordinates on
the brane worldsheet. The functions $X^\mu(\zeta^a)$ determine the
brane worldsheet describing the evolution of the $(D-1)$-dimensional
object (brane) in a bulk $N$-dimensional spacetime. We assume that
$D\le N-1$.  A test brane configuration in an external gravitational
field $g_{\mu\nu}$ can be obtained by solving the equations which
follow from  the Dirac-Nambu-Goto action
\cite{D,NG}
\begin{equation}
S \: = \int d^D\zeta \sqrt{-\mbox{det}\gamma_{ab}}\, ,
\label{action}
\end{equation}
where
\begin{equation}
\gamma_{ab} \: = \: g_{\mu\nu}X^\mu_{,a} X^\nu_{,b}
\label{indmetdef}
\end{equation}
is the $D$-dimensional induced metric on the worldsheet.  Usually the
action  contains the brane tension factor $\mu$. This factor does not
enter into the brane equations. For simplicity we put it equal to 1.

We assume that the brane is static and spherically symmetric,
so that its worldsheet geometry possesses the group  of the
symmetry $O(D-1)$. If $D<N-1$ we choose the brane surface
to obey the equations 
\be\n{equ}
\theta_{N-2}=\ldots=\theta_{D}=\pi/2\, .
\ee
The brane worldsheet with the above symmetry properties is defined by
the function $\theta_{D-1}=\theta(r)$. We shall use 
$\zeta^a=(T,r,\theta_{\hat{i}})$ ($\hat{i}=1, \ldots, n=D-2$) as the
coordinates on the brane. For this parametrization the induced metric
on the brane is
\be\n{mebr}
ds^2=\gamma_{ab}d\zeta^a d\zeta^b=
\ee
\[
-F dT^2+[F^{-1}+r^2 (d\theta/dr)^2] dr^2 + r^2 \sin^2\theta
d\Omega^2_{n}\, , 
\]
and the action \eq{action} reduces to 
\be\n{raction}
S=\Delta T {\cal A}_{n}\int dr {\cal L}\, ,
\ee
\be\n{L}
{\cal L}=r^{n}\, \sin^{n}\theta\, \sqrt{1+F r^2 (d\theta/dr)^2}\, .
\ee
Here $n=D-2$, $\Delta T$ is the interval of time, and
${\cal A}_n=2\pi^{n/2}/\Gamma(n/2)$ is the surface area of a unit
$n$-dimensional sphere. A brane configuration is determined by
solutions of the following Euler-Lagrange equation
\be
{d\over dr}\left( d{\cal L}\over d\theta/dr \right)-{d{\cal L}\over
d\theta}=0\, ,
\ee
which for the Lagrangian \eq{L} is of the form
\be\n{breq}
{d^2\theta\over dr^2}+B_3 \left({d\theta\over dr}\right)^3+B_2 \left({d\theta\over
dr}\right)^2+B_1 {d\theta\over dr}+B_0=0\, ,
\ee
\be
B_0=-{n\cot \theta \over F \, r^2}\hhh
B_1={n+2\over r}+{1\over F}{dF\over dr}\, ,
\ee
\be
B_2=-n\cot \theta\hhh
B_3=r\left[ (n+1)F+ {r\over 2} {dF\over dr}\right]\, .
\ee

For the brane crossing the horizon, the equation \eq{breq} has a
regular singular point at $r=r_0$.  A regular at this point solution
has the following expansion near it
\be
\theta(r)\approx \theta_0 +\theta_0' (r-r_0)+\ldots\hh
\theta_0'={n \cot \theta_0\over 2\kappa r_0^2}\, .
\ee 
This super-critical solution is uniquely defined by the initial value
$\theta_0$.

If the  brane does not cross the horizon, a radius $r$ on the brane
surface reaches its  minimal value $r=r_1>r_0$. For the symmetry
reason it occurs at $\theta=0$. A regular solution of \eq{breq} near
this point has the following behavior
\be\n{ast}
\theta\approx \pm \sqrt{2(n+1)\over B_3(r_1)}\sqrt{r-r_1}+\ldots\, .
\ee
Such a sub-critical solution of is uniquely determined by the value
of the parameter $r_1$.

\section{Far distance solutions}

Consider a brane located in the equatorial plane $\theta=\pi/2$.
Since the bulk metric is invariant under the discrete transformation
$\theta\to \pi-\theta$, the surface $\theta=\pi/2$ is geodesic, and
hence minimal. This means that $\theta=\pi/2$ is a solution of the
test brane equations. This can be also easily checked by using the
equation \eq{breq}. 

Let us consider now a solution which asymptotically approaches
$\theta=\pi/2$. We write this solution in the form
\be\n{asq}
\theta={\pi\over 2}+q(r)\, .
\ee
Assuming that $q$ is small and using \eq{as} one can write the
equation \eq{breq} in the region $r\to \infty$ as
\be
{d^2 q\over dr^2}+{n+2\over r}{dq\over dr}+{n\over r^2} q=0\, .
\ee
For $n>1$ this equation has a solution 
\be\n{asb}
q={p\over r}+{p'\over r^{n}}\, .
\ee
The case $n=1$ is a degenerate one and the corresponding solution
is
\be
q={p+p'\ln r\over r}\, .
\ee
This case was considered in details earlier in \cite{cfl,flc}. In
this paper we focus on the higher dimensional BBH-systems and assume
that $n>1$. In this case the first term in \eq{asb} is the leading
one. The brane surface is asymptotically parallel to the equatorial
plane, and $p=\lim_{r\to\infty}r q$ is the distance of the brane from
it. We shall call this quantity $p$ a {\em shift parameter}.

If the brane  crosses the horizon (a super-critical solution) it is
uniquely determined by the angle $\theta_0$, and the asymptotic data
$\{p,p'\}$  are well defined continuous functions of $\theta_0$. In a
general case, the function $p(\theta_0)$ may be non-monotonic in the
interval $(0,\pi/2)$, so that for two different values of $\theta_0$
one has the same value of the shift parameter $p$.  If the brane does
not cross the horizon (a sub-critical solution) it is uniquely
determined by the minimal radius $r_1$ and the asymptotic data
$\{p,p'\}$  are continuous functions of $r_1$. In a general case it
may also happen that for the same shift parameter $p$ one has two (or
more) solutions, and, for example, one of them is sub-critical and
another super-critical  \cite{cfl,kmmw}.

\begin{figure}[ht]
\epsfig{file=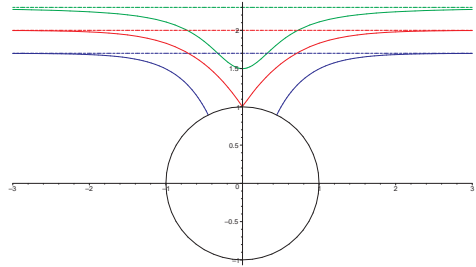, height=3.5cm}
\caption{{\bf A brane interacting with a black hole:} The round circle is
the black hole horizon.   We schematically show the profiles of the 
brane for 3 different cases. The brane which touches the horizon at
one point and has a cone-like profile near it is critical. Two other
solid lines show branes close to the critical one. One of them 
(sub-critical) does not intersect the horizon, while the other 
(super-critical) one crosses the horizon at the spherical surface with
a small radius $R_0$. At far distance the brane surfaces are parallel
to the equatorial plane of the black hole.}
\label{brane}
\end{figure}

We consider sub-critical and super-critical solutions as two
different phases of the BBH-system. Let us consider a continuous
deformation of solutions during which a solution transits from one
phase to another. If we parametrize these solutions by a parameter
$\lambda$, there exists a special value of it, $\lambda=\lambda_*$,
where the phase is changed.  We call the corresponding solution {\em
critical}. ( Figure~\ref{brane} schematically shows the critical and
near critical solutions for this process.) We focus our attention on
the solutions close to the critical one and study the properties of
the BBH-system  during the {\em merger transition} when the solution
changes its phase.

\section{Brane configurations near the horizon}

Near the horizon the coefficient $F$ of the bulk metric has the form
\be
F\approx 2\kappa (r-r_0)+ O((r-r_0)^2)\, ,
\ee
where $\kappa={1\over 2}(dF/dr)|_{r_0}$ is the surface gravity.
To study the near-horizon behavior of the brane it is convenient to
introduce the proper distance coordinate
\be
Z=\int_{r_0}^r {dr\over \sqrt{F}}\, .
\ee
In the vicinity of the horizon one has
\be
r-r_0\approx \kappa Z^2/2\hh
F\approx\kappa^2 Z^2\, .
\ee
We are interested in the case when a `central' part of the brane is
located in the close vicinity of the horizon or crosses it. In the
latter case, we assume that the radius $R_0$ of the surface of the
intersection of the brane with the bulk horizon is much smaller than
the size of the horizon $r_0$. Under these conditions one can
approximate the spaceime close to the bulk black hole horizon by the
Rindler space where the horizon is a $(n+1)$-dimensional plane. 
In this
approximation the metric \eq{met} takes the form
\be
dS^2=-\kappa^2 Z^2 dT^2+dZ^2+dL_{N-2}^2\, ,
\ee
where $dL_{N-2}^2$ is the metric of $(N-2)$-dimensional Euclidean
space $\E^{N-2}$. We write $dL_{N-2}^2$ in the form
\be
dL_{N-2}^2=dY^2_{D}+\ldots + dY^2_{N-2}+dR^2+R^2d\Omega^2_{D-2}\, ,
\ee
and choose the Cartesian coordinates $Y$ so that the equation \eq{equ}
takes the form $Y_{D}=\ldots = Y_{N-2}=0$, while the brane equation is 
\be\n{const}
F(Z,R)=0\, .
\ee
We write a solution of this equation in a parametric form
\be
Z=Z(\lambda)\hh R=R(\lambda)\, .
\ee
Then the induced metric on the brane is
\ba\n{hmet}
ds^2&=& -\kappa^2 Z^2 dT^2 \non\\
&+&[(dZ/d\lambda)^2+(dR/ d\lambda)^2] d\lambda^2
+R^2 d\Omega^2_{n}\, .
\ea
Here, as earlier, $n=D-2$.
The action \eq{action} for this induced metric takes the form
\be
S=\kappa \Delta T {\cal A}_{n} {\cal S}\, ,
\ee
\be
{\cal S}=\int d\lambda  Z R^n \sqrt{
(dZ/d\lambda)^2+(dR/d\lambda)^2}\,  .
\ee
This action is evidently invariant under a reparametrization
$\lambda\to\tilde{\lambda}(\lambda)$. In the regions where either $Z$
or $R$ is a monotonic function of $\lambda$, these functions
themselves can be used as parameters. As a result, one obtains two
other forms of the action which are equivalent to ${\cal S}$
\be
{\cal S}=\int dZ {\cal L}_R=\int dR {\cal L}_Z\, ,
\ee 
where
\be
{\cal L}_R=Z R^{n} \sqrt{1+{R'}^2}\hh
{\cal L}_Z=Z R^{n} \sqrt{1+{\dot{Z}}^2}\, .
\ee
Here the prime means the derivative with respects to $Z$, while the
dot stands for the derivative with respect to $R$.
The corresponding Euler-Lagrange equations are
\be\n{eqR}
Z R R'' +(R R'-nZ)(1+{R'}^2)=0\, .
\ee
\be\n{eqZ}
R Z \ddot{Z} +(nZ \dot{Z}-R)(1+{\dot{Z}}^2)=0\, .
\ee
It is easy to check that the form of the equation \eq{eqR} is
invariant under the following transformation
\be\n{scR}
R(Z)=k\tilde{R}(\tilde{Z})\hh
Z=k\tilde{Z}\, .
\ee
Similarly, the transformation
\be\n{scZ}
Z(R)=k\tilde{Z}(\tilde{R})\hh
R=k\tilde{R}
\ee
preserves the form of \eq{eqZ}.

To obtain the boundary conditions to the equations \eq{eqR} and
\eq{eqZ} we require that the induced metric on the brane is regular.
This implies that the curvature invariants are regular as well. Let
us consider the scalar curvature of the induced metric, which we
denote by ${\cal R}$. For the brane parametrization $R=R(Z)$ the
induced metric is
\be
ds^2= -\kappa^2 Z^2 dT^2 +[1+{R'}^2] dZ^2+R^2 d\Omega^2_{n}\, ,
\ee
and the corresponding scalar curvature takes the form
\be
{\cal R}={\cal R}^{(2)}+{n(n-1)\over R^2}-2n{\Box^{(2)}R\over
R}-n(n-1){(\nabla R)^2\over R^2}\, .
\ee
The quantities  in the right-hand side of this relation are
calculated for the two-dimensional metric which is the metric of the
$(T-Z)$-sector of the \eq{hmet}. In particular, ${\cal R}^{(2)}$ is the
two-dimensional  curvature of this metric. Calculating ${\cal R}$ and
using the brane equation \eq{eqR} to exclude the second derivatives
$d^2 R/dZ^2$ one gets
\be
{\cal R}={2R^2 {R'}^2-6n ZR R'+n(3n-1)Z^2\over Z^2 R^2 (1+{R'}^2)}\, .
\ee

Simple analysis shows that if the brane crosses the horizon of the
bulk black hole, then the regularity of ${\cal R}$ on the horizon
requires
\be\n{bcR}
R|_{Z=0}=R_0\hh \left.{dR\over dZ}\right|_{Z=0}=0\, .
\ee
By using \eq{eqR} one obtains that for this solution near the horizon
one has
\be\n{nhR}
R=R_0+{n Z^2\over 4 R_0}+\ldots \,. 
\ee 

For a brane which does not cross the horizon one has $Z(R=0)>0$, and
at this point one has $R'\to \infty$. This gives the following boundary
conditions
\be\n{bcZ}
Z|_{R=0}=Z_0\hh \left. {dZ\over dR}\right|_{R=0}=0\, . 
\ee
These conditions can also be obtained from the regularity of ${\cal
R}$ in the parametrization
$Z=Z(R)$. Using \eq{eqZ} one gets
\be\n{nhZ}
Z=Z_0+ {R^2\over 2(n+1)Z_0} +\ldots \,.
\ee
One can check that when the condition of regularity of the
Ricci scalar is satisfied, the other curvature invariants are also
finite.

A solution  near the horizon  (either  \eq{bcR} or \eq{bcZ})
determines uniquely a global solution of the brane equation
\eq{breq}. Thus for a given small value of $R_0$ ($Z_0$) the
asymptotic data of the corresponding solution, $\{ p,p'\}$, are
uniquely determined and are continuous functions of $R_0$ ($Z_0$).

\section{Critical solutions as attractors}

The equations \eq{eqR}-\eq{eqZ} have a simple solution 
\be\n{crit}
R=\sqrt{n} Z
\ee
which plays a special role. We call it a {\em critical solution}. It
describes a critical brane which touches the horizon of the bulk
black hole at one point, $Z=R=0$. The critical solution separates two
different families of solutions (phases), super-critical and
sub-critical.  Let us show that the critical solution is an {\em
attractor}, and solutions of the both families are attracted to the
critical solution asymptotically.

To do this, we, following \cite{flc}, introduce new variables
\be
x=R'\hhh
y=Z^{-1}RR'\hhh
ds=dZ/(yZ)\, .
\ee
In these variables the equation \eq{eqR} can be written in the form of
the first order regular autonomous system
\ba\n{sys1}
{dx\over ds}&=&x(n-y)(1+x^2)\, ,\\
\n{sys2}
{dy\over ds}&=&y[n-2y+x^2(n+1-y)]\, .
\ea

\begin{figure}[ht]
\epsfig{file=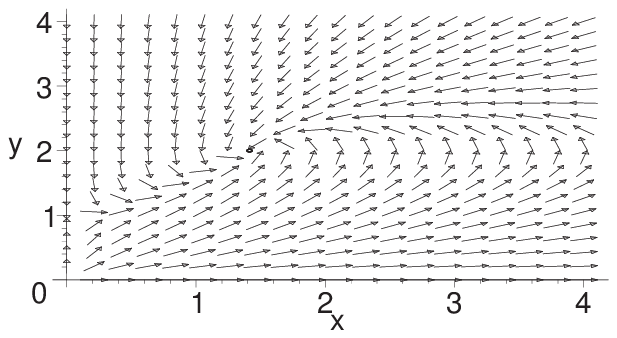, height=4cm,width=4cm}
\hfill
\epsfig{file=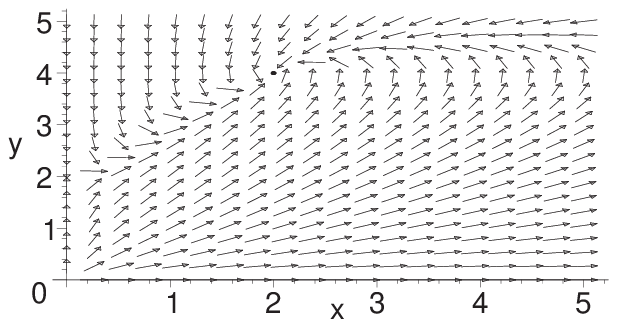, height=4cm,width=4cm}
\caption{Phase portraits of the system \eq{sys1}-\eq{sys2} for $n=2$
(left plot) and for $n=4$ (right plot).}
\label{PhP}
\end{figure}

The critical points on the phase space $(x,y)$ are the points where
the right-hand side of the both equations \eq{sys1} and \eq{sys2}
vanish simultaneously. These critical points are: a node, $(0,0)$; a
saddle point, $(0,n/2)$, and two focus points, $(\pm\sqrt{n},n)$. 
For different $n$ the phase portrait are similar. The
Figure~\ref{PhP} shows the phase portraits for $n=2$ and $n=4$, which
have the focus points at $(\sqrt{2},2)$ and $(2,4)$, respectively.
The focus points are attractors, and they correspond to the critical
solutions $R=\sqrt{n}Z$.

\section{Near-critical solutions in the vicinity of the horizon}

Let us consider now solutions which are close to the critical one
\be\n{ncrit}
R=\sqrt{n}Z+\rho(Z)\, .
\ee
Substituting this expression into \eq{eqR}, and keeping linear in $\rho$
terms we obtain
\be\n{linR}
Z^2 \rho''+(n+1)Z \rho'+(n+1)\rho=0\, .
\ee
A solution of this equation is of the form 
\be\n{p}
\rho=Z^{\beta}\hh
\beta_{\pm}={1\over 2}(-n\pm \sqrt{n^2-4n-4})\, .
\ee
The parameter $n$ is a positive integer. For $n\le 4<2(1+\sqrt{2})$ the
exponent $\beta$ is complex, while for $n\ge 5$ it is real.

Let us consider first the case of complex $\beta$, that is when the
number of the brane spaceime dimensions is $D\le 6$. We write a
near-critical solution \eq{ncrit} in the form
\be\n{ncs}
R=\sqrt{n}Z+Z^{-n/2}\Re(C Z^{i\alpha})\, ,
\ee
where $C$ is a complex number and $\alpha={1\over 2}\sqrt{4+4n-n^2}$.
The initial data $R(Z=0)=R_0$ determines the factor $C$. We denote the
corresponding values of it by $C(R_0)$. Under the scaling transformation
\eq{scR} the surface $Z=0$ is invariant, while $R_0$ is transformed
into $\tilde{R}_0=k^{-1}R_0$. Using \eq{ncs} we obtain
\be\n{scC}
C(kR_0)=k^{-1-n/2+i\alpha}C(R_0)\, .
\ee

For $D\ge 7$ the both exponents $\beta_{\pm}$ are real and one has
$\beta_+\beta_-=4(n-1)$, so that both $\beta_{\pm}$ are negative. The
critical solution \eq{crit} is again the attractor and one has
\be
R=\sqrt{n}Z+C_+Z^{\beta_+}+C_-Z^{\beta_-}\, .
\ee
Using again the arguments based on the scaling properties of the
equation \eq{eqR} one can conclude that $C_{\pm}$ considered as
functions of $R_0$ obey the properties
\be\n{scCa}
C_{\pm}(kR_0)=k^{\beta_{\pm}-1}C_{\pm}(R_0)\, .
\ee

Similarly, one can consider near-critical solutions of the equation
\eq{eqZ}. Let us write
\be
Z={1\over \sqrt{n}}R+\zeta(R)\, .
\ee
Then keeping linear in $\zeta$ terms one obtains
\be\n{linZ}
R^2\ddot{\zeta}+(n+1)R\dot{\zeta}+(n+1)\zeta=0\, .
\ee
It has the same form and the same coefficients as the equation
\eq{linR}. For this reason, as it can be expected, the critical
dimension and the scaling property \eq{scR} are the same for the both
linearized equations.

\section{Scaling and self similarity}

Since near the horizon the equation \eq{breq} reduces to \eq{eqR},
one can uniquely define a special global solution of \eq{breq} which
near the horizon reduces to the critical solution \eq{crit}. This
critical solution near the horizon is of the form \cite{rem}
\be
\theta\approx \sqrt{ {2n(r-r_0)\over \kappa r_0^2}}\, .
\ee
Being traced to infinity, the critical solution determines uniquely
the asymptotic data $\{p,p'\}$ which we denote by  $\{p_*,p'_*\}$,
respectively. If the function $F(r)$ in the metric is given one can
find $\{p_*,p'_*\}$, for example, by solving the equation \eq{breq}
numerically.

Consider a super-critical solution. For this solution the radius
$R_0$ of the horizon of the brane black hole is connected with
$\theta_0$ as follows $R_0=r_0\sin\theta_0$. For small $\theta_0$ one
has $R_0=r_0\theta_0$.  We denote by $\{ p,p'\}$ the asymptotic data
for a solution with a given $R_0$. For the critical solution,
$R_0=0$, the asymptotic data are $\{ p_*,p'_*\}$. We  also denote
$\Delta p=\sqrt{(p-p_*)^2+(p'-p'_*)^2}$.  Similarly, one can define
$\Delta p$ for sub-critical solutions. In the both cases for the
critical solution, and only for it, $\Delta p=0$ .  One can use
$\Delta p$ as a measure indicating how close a solution is to the
critical one.  Our goal is to demonstrate that $R-0$ (or $Z_0$)
considered as a function of $\Delta p$ has a self-similar behavior. We
shall show that there exists a critical dimension, $D_*$, such that
for $D> D_*$  this symmetry is continuous, while for $D\le D_*$ it is
discrete. This critical dimension for the BBH-system is $D_*=6$.
It should be emphasized that in the definition   one can use any
positive definite quadratic form of $p-p_*$ and $p'-p'_*$  of
$\Delta p$ instead of $(p-p_*)^2+(p'-p'_*)^2$. It can be
checked that this does not change the result.

\subsection{$D\le 6$ case}

Since in the  solution \eq{crit} is the attractor, both  sub-critical
and super-critical solutions which are close to the critical solution
in the vicinity of the horizon remain close to it everywhere. This
implies that the difference between the near-critical and critical
solutions is small and it can be obtained by solving  a linear
equation. For this equation both $(p-p_*, p'-p'_*)$ and the complex
coefficient $C$ in \eq{ncs} are uniquely defined by $R_0$, so that
there exists a linear relation between these coefficients
\be
C=a(p-p_*)+b(p'-p'_*)\, ,\ \bar{C}=\bar{a}(p-p_*)+\bar{b}(p'-p'_*)\, ,
\ee
where $a$ and $b$ are complex numbers.
Substituting $\{p,p'\}$ obtained from these equations into the
definition of $\Delta p$ one gets
\ba
(\Delta p)^2&=&P^2\left[ C \bar{C}-\bar{{\cal A}} C^2-{\cal
A}{\bar{C}}^2\right]\, ,\\
{\cal A}&=&|{\cal A}|e^{iA}={a^2+b^2\over 2(|a|^2+|b|^2)}\, ,\\
P^2&=&-{2(|a|^2+|b|^2)\over (a\bar{b}-\bar{a}b)^2}\, .
\ea
It is easy to show that $P^2$ is non-negative, and 
\be\n{ineq}
|{\cal A}|\le 1/2\, .
\ee
In the latter relation the equality takes place when
$a/\bar{a}=b/\bar{b}$. This is a degenerate case and by a small
change of the bulk metric this condition will be violated. Thus we
assume that $|{\cal A}|< 1/2$.

We denote by $\Delta \tilde{p}$ and $\tilde{C}$  quantities which
correspond to the solution with $R=\tilde{R}_0$. Then one has
\be
{(\Delta p)^2\over (\Delta \tilde{p})^2}={ C \bar{C}-\bar{{\cal A}} C^2-{\cal
A}{\bar{C}}^2 \over  \tilde{C} \bar{\tilde{C}}-\bar{{\cal A}}
\tilde{C}^2-{\cal A}{\bar{\tilde{C}}}^2}\, .
\ee

The scaling transformation \eq{scR} maps $Z=0$ onto itself, and hence
it preserves the position of the horizon, while it changes the value
of $R_0$. Let us choose $k=R_0/\tilde{R}_0$, then, using \eq{scC},
one has
\be
C=(R_0/\tilde{R}_0)^{(n+2)/2-i\alpha}\tilde{C}\hh \gamma=1+n/2\, .
\ee
Let us denote
\be
{\cal B}={\cal A}{\bar{\tilde{C}}\over \tilde{C}}=|{\cal A}|e^{iB}\, .
\ee
Then one has
\be
{(\Delta p)^2\over (\Delta \tilde{p})^2}
={R_0^{n+2}\, [1-2|{\cal A}|\cos(2\alpha \ln R_0 +B')]\over 
\tilde{R}_0^{n+2}\, [1-2|{\cal A}|\cos(B)]}\, ,
\ee
where $B'=B-2\alpha \ln \tilde{R}_0$. It follows from this relation
that
\ba\n{scal}
\ln(\Delta p)&=&{n+2\over 2} \ln R_0+H \non \\
&+&{1\over 2} \ln[1-2|{\cal A}|\cos(2\alpha \ln
R_0 +B')]\, ,
\ea
where $H$ and $B'$ do not depend on $R_0$.

For small $R_0$ the leading part of this equation gives
\be
\ln R_0\sim {2\over n+2}\ln (\Delta p)\, .
\ee
Using this relation and iterating \eq{scal} one obtains
\be\n{SC}
\ln R_0\sim \gamma \ln (\Delta p) +f(\ln (\Delta p))+C\, .
\ee
Here $C$ is a constant,
\be\n{g}
\gamma={2\over n+2}\, ,
\ee
and $f(z)$ is a periodic function of $z$, $f(z+\omega)=f(z)$ with the
period $\omega$
\be\n{w}
\omega={\pi (n+2)\over \sqrt{4+4n-n^2}}\, .
\ee
For $n=1$ relations \eq{g} and \eq{w} reproduce the result obtained in
\cite{cfl,flc}.

\subsection{$D> 6$ case}

In this case the real coefficients $C_{\pm}$ are connected with the
asymptotic data $\{ p,p'\}$ as follows
\be
p-p_*=a_+ C_{+}+a_- C_{-}\hhh
p'-p'_*=a'_+ C_{+}+a'_- C_{-}\, ,
\ee
and one has
\ba\n{Dp7}
(\Delta p)^2&=&(a_+^2+{a'_+}^2)C_+^2 + (a_-^2+{a'_-}^2)C_-^2 \non \\
&+&2(a_+ a_- +a'_+ a'_-) C_+C_-  \, .
\ea
The equation \eq{scCa} implies
\be
C_{\pm}=\left({R_0\over
\tilde{R}_0}\right)^{1-\beta_{\pm}}\tilde{C}_{\pm}\, .
\ee
Since $a_+^2+{a'_+}^2$ does not vanish and $1-\beta_{\pm}>0$,
$1-\beta_+<1-\beta_-$, the first term in the right hand side of the
relation \eq{Dp7} is the leading one for small $R_0$.  $(\Delta p)^2$
as a function of $R_0$ has the following asymptotic form for small
$R_0$
\be
\ln (\Delta p)^2\sim 2(1-\beta_+) \ln R_0\, .
\ee
Or, equivalently, one has
\be\n{sch}
\ln R_0\sim \gamma \ln (\Delta p)\hhh
\gamma={n+2+\sqrt{n^2-4n-4}\over 4(n+1)}\, .
\ee
Hence for the number of dimensions higher than the critical one 
has the scaling law \eq{sch}. The sub-leading oscillations are
absent and the symmetry is continuous.

The relations \eq{SC}-\eq{w} and \eq{sch} demonstrate the scaling law
for the super-critical solutions close to the critical one. It is
easy to check that similar relations are valid for sub-critical
solutions as well. 

\section{Merger transitions: Inside brane story}

Let us imagine that there is an observer `living' within the brane
who is not aware of the existence of the bulk extra-dimensions. Let
us discuss what happens from his or her point of view when the
BBH-system changes its phase. As earlier we consider a one-parameter
family and assume that the merger transition occurs at the special
value of the parameter $\lambda=\lambda_*$. From the point of view of
the brane observer this family describes a process in which a brane
black hole is either created or destroyed. The surface gravity of the
brane black hole always coincides with the surface gravity $\kappa$ of
the bulk black hole, and hence it remains constant during this
process.  For the description of the transition it is convenient to
make the Wick's rotation $T\to i\tau$ and to choose the Euclidean
time $\tau$ to have the period $\kappa/(2\pi)$. In such (`canonical'
by the terminology adopted in \cite{kol}) approach the Euclidean
topology of the BBH-system in two different phases is different.
Namely, for family of sub-critical solutions the topology is
$S^1\times R^{D-1}$, while for the super-critical solutions the
topology is $R^2\times S^{D-2}$. Following \cite{kol}, we call
such a topology change transition the {\em merger transition}. 

The internal geometry of the brane is determined by its embedding into
the bulk spacetime. But if the brane observer does not know about the
existence of extra-dimensions, he/she may try to  use
the `standard' $D$-dimensional Einstein equations in order to interpret
the observed brane geometry. Such an observer would arrive to
the conclusion that the brane spacetime is not empty and there exists a
distribution of the matter in it. By using the $D$-dimensional
Einstein equations 
\be
G_{ab}=R_{ab}-{1\over 2}\gamma_{ab}R={\cal T}_{ab}\hh
{\cal T}_{ab}=8\pi T_{ab}
\ee
one can obtain the corresponding effective stress-energy tensor $T_{ab}$.
(We use units in which the $D$-dimensional gravitational coupling
constant is $1$.) To calculate $G_{ab}$ we write the induced metric on
the brane \eq{mebr} in the form ($A,B=(0,1)$)
\be\n{im}
ds^2=\gamma_{ab}d\zeta^a d\zeta^b=h_{AB}dy^A dy^B+\rho^2(y)d\Omega_n^2\, .
\ee
The symmetry of the metric \eq{im} implies that ${\cal T}_{a}^{b}$
has the following non-vanishing components 
\be
{\cal T}^{b}_{a}= \left( 
\begin{array}{ccccc}
{\cal T}^0_0 & {\cal T}^0_1 & 0 & \ldots & 0 \\
{\cal T}^0_1 & {\cal T}^1_1 & 0 & \ldots & 0 \\
0 & 0 & \hat{\cal T} & \ldots & 0 \\
\ldots & \ldots &\ldots &\ldots &\ldots \\
0 & 0 & 0 & \ldots & \hat{\cal T} 
\end{array}
\right) .
\ee
The calculations give
\ba\n{set}
{\cal T}_{AB}&=&n
\left[-{\rho_{:AB}\over \rho}+ \left({\Box^{(2)} \rho\over \rho} 
+\displaystyle\frac{n-1}{2\rho^2}
{\cal P}\right) h_{AB} \right]
\, ,\nonumber \\
{}\\
\hat{{\cal T}}&=&
(n-1) \left[
{\Box^{(2)}\rho \over\rho} +\displaystyle\frac{n-2}{2\rho^2} {\cal P}\right]
-{1\over 2}\,{\cal R}^{(2)},\nonumber
\ea
where ${\cal P}=(\nabla \rho)^2-1$.

For the super-critical solutions, substituting \eq{nhR} into \eq{set} one
obtains at the horizon ($Z=0$)
\be\n{TR}
{\cal T}_T^T={\cal T}_Z^Z={n\over 2R_0^2}\hh
\hat{{\cal T}}={n^2+2n-4\over 4R_0^2}\, .
\ee
Similarly, for the sub-critical solutions, substituting \eq{nhZ} into
\eq{set} one obtains at the top of the brane ($R=0$)
\be\n{TZ}
{\cal T}_T^T=-{n\over 2(n+1)Z_0^2}
\hhh
{\cal T}_R^R=\hat{{\cal T}}={n(n+3)\over 2(n+1)^2 Z_0^2}\, ,
\ee
The equations \eq{TR} and \eq{TZ} show that in the both phases the
tensor $T_a^b$ calculated at the horizon respects the symmetries of
the induced metric.

\section{Discussion}

We discussed merger transitions in the brane--black-hole system. We
would like to emphasize that there exists a striking similarity
between this phenomenon and the merger transitions in the
black-hole--black-string system \cite{kol,kol1}. Namely, close to the
horizon the near critical solutions of the BBH-system has the same
`double cone' structure as the solutions for the merger transition of
the caged black holes. The  equations defining the near critical
solutions, following from the Einstein action for the latter system,
and the equations for the BBH-system are very similar. In both cases
there is a critical dimension of the spacetime where  the scaling
parameter $\alpha$ becomes real. For $D=3$, the parameter $\alpha$ is
the same in the both cases.

Kol \cite{kol} discussed a possible relation between merger
transitions and the Choptuik phenomena. This similarity with the
critical collapse takes place in the BBH-system as well.  To
demonstrate this let us consider a slow evolution of the BBH-system
in time which starts with no-brane-black-hole, so that a black hole
is formed as a result of time evolution. As earlier, one has a
one-parameter family of quasi-static configurations, and at the
moment of brane-black-hole formation the asymptotic data $(\{p,p'\})$
reaches the critical value $(\{p_*,p_*'\})$. The relations \eq{TZ}
show that the curvature at the center of the brane is growing until
it formally becomes infinite at the moment of the black hole
formation. For $D\le 6$ the growing of the curvature is periodically
modulated as it happens in the critical collapse case \cite{GD,GCD}.
For the number of dimensions higher than the critical one (for $D>
6$) this oscillatory behavior disappears. The scaling laws, relating
the size $R_0$ of the formed brane-black-hole with $\Delta p$, also
has the same scaling and self-similar behavior as in the critical
collapse case \cite{chop}.

In our analysis we focused on the static brane solutions and
neglected the effects connected with the brane tension. Let us discuss
what happens when the brane is moving. If the brane approaches the
bulk black hole the BBH will be created. In the inverse case when the
brane, initially crossing the bulk black hole horizon, is moving away
from the black hole, the BBH can disappear. In a general case, for
finite velocity of the brane one cannot use the adiabatic
approximation and describe the brane motion as a set of static
configurations. To study dynamics of this process one needs to
include a time variable explicitly and  to solve a 2-dimensional
problem. This can be done numerically. In many aspects this problem 
is similar to (but more complicated than) numerical solving of
equations for moving cosmic strings interacting with the bulk black
hole \cite{SFV,SF}. Let us emphasize that the dynamics
of the process of BBH disappearance is not a time reversed version of
the BBH formation process. 

The reason of this time asymmetry is connected with the presence of
the bulk black hole. Consider a bulk black hole formed as a result of
the gravitational collapse of some matter. By definition, the black
hole is a region of the bulk spacetime causally disconnected from the
future null infinity. A solution obtained by inversion of
the direction of time $t\to -t$  describes a completely different
physical process, an expansion of the matter from the white black
hole (see e.g. \cite{FN}). Physical processes in the spacetime of a black hole
obey natural condition of regularity at the future event horizon. To
study such processes in the spacetime of a black hole long time after
its formation one can "forget" about the details of the gravitational
collapse and to use the eternal black hole approximation  imposing
the same regularity conditions at the future event horizon. For a static
solutions this regularity conditions imply its regularity at the past
horizon of the eternal black hole. For time-dependent configurations
in a general case  this is not true. 

Interaction of a moving brane with the black hole was studied
numerically in \cite{FT,FPSTa,FPSTb}. A main motivation for this
study is connected with the following question: under which
conditions a black hole moving with some velocity can escape from the
brane \cite{FSa,FSb}. By solving numerically test brane equations in
a spacetime of a static black hole it was shown \cite{FT} that in the
considered cases a moving brane is bend and eventually the radius of the
pinched part goes to zero. This indicates that the extraction of the
test brane from the bulk black hole is accompanied by its
reconnection \cite{FT}. After the reconnection a peace of the brane near
the horizon ("a baby brane") is absorbed by the black hole, while the
other part,  located outside of the black hole ("a mother brane"), 
keeps moving away from the black hole. In such a process a brane
observer registers that the BHH disappears at the moment of
the reconnection. 

In the limit of small velocity of the brane, effects connected with
the brane tension cannot be ignored and they can significantly change
the dynamics of the system. Consider a $D$-dimensional brane with the
tension $\sigma$ in the spacetime of $N$-dimensional black hole with
the gravitational radius $r_0$. For these parameters one can
construct the following dimensionless combination
$\epsilon=G^{N}\sigma/r_0^{N-D-2}$, where $G^{N}$ is the Newtonian
coupling constant in the bulk spacetime. It is argued in
\cite{FT,FPSTb} that the escape velocity for such a system is of the
order of $\epsilon^{1/2}$.

In the above discussion we assumed that the brane is infinitely thin.
In the field theory with the spontaneous symmetry breaking the
brane-like objects (e.g domain walls) arise as a solution of
non-linear equations, and they have finite thickness. Static thick
domain walls interacting with a black hole were studied in
\cite{MYIIN,MIIK,Rog,MRa,MRb}. 

One can expect that  for an infinitely thin brane in the process of
its reconnection, during which the BBH disappears,the curvature of
the induced metric infinitely grows as it happens for the critical
solutions discussed above. Under these conditions one cannot neglect
the finite thickness effect. One can expect that formal singularities
in the infinitely thin brane description would disappear when one
uses a (more fundamental) field theory description.  In this
connection the recent numerical calculations performed in \cite{FPST}
are very stimulating. It is interesting to analyze  in more details
the dynamics of the destruction of BBHs in the framework of the field
theory models. For a finite-thickness brane a choice of the surface,
which represents it, is not unique. Thus for the thick brane the
meaning of the induced geometry is not well defined. If the gravity
is emergent phenomenon the merger and topology change transitions in
the physical spacetime might have similar features with those of the
discussed toy model. Namely in the vicinity of these transitions one
cannot any more use the metric for the description of the details of
the transition. Instead a more detailed microscopic description in
terms of constituents (e.g. strings) is required. The toy model
proposed in this paper and its field theory analogues might be useful
for modeling these transitions.

\noindent  

\section*{Acknowledgments}  
\noindent  
The author benefited a lot from the discussions of merger transitions
with Barak Kol. He is also grateful to Evgeny Sorkin and David Mateos
for useful remarks. The research was supported in part by US National
Science Foundation under Grant No. PHY99-07949, by the Natural
Sciences and Engineering Research Council of Canada and by the Killam
Trust.


\end{document}